\documentclass[aps,
               prb,
               reprint,
               superscriptaddress,
               floatfix,
              ]{revtex4-1}

\usepackage[english]{babel}

\usepackage[tbtags]{amsmath}

\usepackage{bm}              % bold greek letters

\usepackage{graphicx}

\usepackage{dcolumn}    %% align table column on decimal point(s)
\newcommand{\etal}{\textit{et~al.\ }}    % and others

\begin{document}

\title{Theoretical study of magnetic domain walls through a cobalt nanocontact}

\author{L\'aszl\'o Balogh}
%\email[]{blaci947@yahoo.co.uk}
\affiliation{Department of Theoretical Physics, Budapest University of Technology and Economics, H-1111 Budapest, Hungary}
\affiliation{Condensed Matter Research Group of the Hungarian Academy of Sciences, Budapest University of Technology and Economics, H-1111 Budapest, Hungary}

\author{Kriszti\'an Palot\'as}
%\email[]{palotas@phy.bme.hu}
\affiliation{Department of Theoretical Physics, Budapest University of Technology and Economics, H-1111 Budapest, Hungary}

\author{L\'aszl\'o Udvardi}
\email[Corresponding author. ]{udvardi@phy.bme.hu}
\affiliation{Department of Theoretical Physics, Budapest University of Technology and Economics, H-1111 Budapest, Hungary}
\affiliation{Condensed Matter Research Group of the Hungarian Academy of Sciences, Budapest University of Technology and Economics, H-1111 Budapest, Hungary}

\author{L\'aszl\'o Szunyogh}
%\email[]{szunyogh@phy.bme.hu}
\affiliation{Department of Theoretical Physics, Budapest University of Technology and Economics, H-1111 Budapest, Hungary}
\affiliation{Condensed Matter Research Group of the Hungarian Academy of Sciences, Budapest University of Technology and Economics, H-1111 Budapest, Hungary}

\author{Ulrich Nowak}
\affiliation{Department of Physics, University of Konstanz, 78457 Konstanz, Germany}
\date{\today}

\begin{abstract}
     To calculate the magnetic ground state of nanoparticles we present a self-consistent first principles method
     in terms of a fully relativistic embedded cluster multiple scattering Green's function technique.
     Based on the derivatives of the band energy, a Newton-Raphson algorithm is used to
     find the ground state configuration. The method is applied to a cobalt nanocontact that turned out to show
     a cycloidal domain wall configuration between oppositely magnetized leads.
     We found that a wall of cycloidal spin-structure is about 30\,meV lower in energy than the one of helical spin-structure.
     A detailed analysis revealed that the uniaxial on-site anisotropy of the central atom is
     mainly responsible to this energy difference.
     This high uniaxial anisotropy energy is accompanied by a huge enhancement and anisotropy of the orbital magnetic moment
     of the central atom. 
     By varying the magnetic orientation at the central atom, we
     identified the term related to exchange couplings (Weiss-field term), various on-site anisotropy terms,
     and also those due to higher order spin-interactions.
\end{abstract}

\maketitle

\section{Introduction}

As magnetic storage devices approach a physical limit of a single atom,
the investigation of nanoclusters has become one of the most important subjects in
magnetism. Recent developments in nanotechnology permit the construction of clusters with
well-controlled structures and enable the measurement of various magnetic properties
at the atomic scale. Probing the Kondo resonance in terms of low temperature
scanning tunneling spectroscopy Heinrich \etal \cite{Heinrich2004} determined the spin-flip
energy of a single manganese atom on a nonmagnetic substrate, while
Wahl \etal \cite{Wahl2007} were able to estimate the exchange coupling
between Co atoms on a Cu(001) surface.
Atomic scale contacts can be fabricated by using electromigrated break
junctions where the size of a macroscopic contact between two leads can be reduced
down to a single atom.
N\'eel \etal \cite{Neel2007} studied the transition from the tunneling to the contact regime
by moving the STM tip closer to the surface adatom, and an enhanced Kondo temperature was found.
In conjunction with the Kondo effect, Calvo \etal \cite{Calvo2009}
found a Fano resonance for ferromagnetic point contacts
indicating that the reduced coordination can dramatically effect
the magnetic behavior of nanoclusters.

Experiments on atomic-sized contacts of ferromagnetic metals generated by mechanically controllable break junction (MCBJ)
revealed magnetoresistance (MR) effects of unprecedented size.~\cite{Chopra2002,Viret2002,Viret2006}
There are various mechanisms to this huge MR discussed in the literature: depending, e.g., 
on the micromagnetic order of the sample controlled by
the size of the applied field, atomically enhanced anisotropic MR (AAMR), giant MR (GMR),
tunnel MR (TMR) or ballistic MR (BMR) effects can be established.~\cite{Egle2010}
In particular, based on ab initio calculations, the AAMR has been shown to emerge in
wire-like transition metal nanocontacts and related to the giant orbital moment
formed at the central atom.~\cite{Autes2008}

Ab-initio calculations on magnetic nanostructures are useful for a clear
interpretation of experimental results and to attain better understanding of the
underlying physical phenomena. Several methods to determine complex magnetic
ground states of nanoparticles from first principles are based on a fully
unconstrained local spin-density approximation (LSDA) implemented
within the full-potential linearized augmented plane-wave (FLAPW) method\cite{Kurz2001}
or the projector augmented-wave (PAW) method.\cite{Hobbs2000}
Unconstrained non-collinear magnetic calculations are also performed within
a tight-binding approach,\cite{Robles2006}
using the tight-binding linearized muffin-tin
orbital (TB-LMTO) method~\cite{Bergman2007-JPCM,Bergman2007-PRB} or the
Korringa-Kohn-Rostoker (KKR) method.\cite{Yavorsky2006}
Spin-orbit coupling (SOC) has an important role in the formation of different magnetic
states via magnetocrystalline anisotropy and Dzyaloshinsky-Moriya (DM)
interactions.~\cite{Bode2007} SOC is usually treated as perturbation
or by directly solving the Dirac equation. The latter concept is applied in
studies relying on ab-initio spin-dynamics in terms of a
constrained LSDA by means of a fully relativistic KKR method.~\cite{Ujfalussy2004,Lazarovits2004,Stocks2007}

In bulk ferromagnets the formation of a domain wall is governed by a
competition between the exchange and anisotropy energies~\cite{Bloch1932} and
the typical interface between the magnetic domains is the Bloch wall where
the magnetization remains perpendicular to the axis of the wall.
In thin films with easy plane anisotropy,
a N\'eel wall is formed with atomic magnetic
moments lying in the plane of the film, however, DM interactions can give rise
to domain walls with out-of-plane magnetization and well-defined rotational sense.~\cite{Heide-dissertation,Heide2008}
In a geometrically constrained system the structure of a
domain wall is mainly determined by the geometry irrespective of the exchange and
anisotropy energies.\cite{Bruno1999}
Thermal effects play an additional role and can lead to new types of domain walls beyond the usual restriction
of constant magnetization magnitude.\cite{Kazantseva2005}
However, for a deeper understanding of the magnetic properties of nanocontacts,
models based on first princples calulations are of pronounced importance.

In the present work, a domain wall through a point-contact between (001) surfaces of fcc
Co is studied, where the magnetizations are aligned in the (110) and the
($\overline{1}\overline{1}$0) directions in the leads.
It should be noted that Co exhibits a hcp structure in bulk, however, as thin film it
often displays a fcc-related geometry.
We apply a fully relativistic embedded cluster Green's function
technique based on the KKR method (EC-KKR).\cite{Lazarovits2002}
Using gradients and second derivatives of the band energy related to the
transverse magnetization, a self-consistent Newton-Raphson method
is developed to find the ground state configuration of the domain wall.
An enhancement of the magnetic anisotropy energy has been
established theoretically in atomic scale junctions even for elements that are nonmagnetic in bulk.~\cite{Thiess2010}
In agreement with this finding, our results reveal that the central atom with the lowest coordination number has the
main contribution to the magnetic anisotropy of the contact.
To highlight the relationship between the obtained cycloidal domain wall configuration
and the magnetic anisotropy, the orientational dependence of the band energy of the point-contact is
analyzed in details.

\section{Computational details}

\begin{figure}[!tb]
\includegraphics[width=\columnwidth]{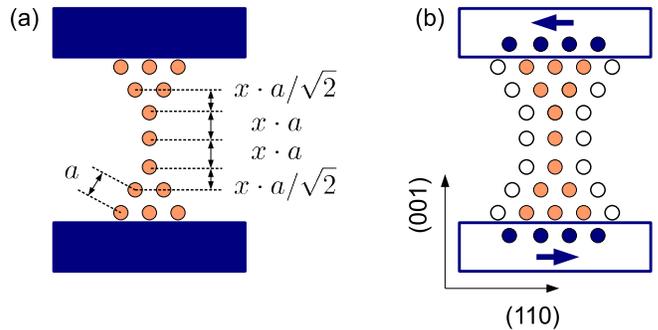}%
\caption{\label{fig:contact}
  (Color online)
  (a)~The geometry of the contact viewed from the $(1\overline{1}0)$ direction.
  The leads are depicted as dark (blue) rectangles, the cobalt atoms forming the contact
  are represented by gray (orange) circles,
  $a$ denotes the nearest neighbor distance in the fcc structure.
  The length of the contact is tuned via $x=$ 0.85, 0.90, 0.95, 1.00, 1.05, 1.10, and 1.15.
  Note that only the marked distances were scaled.
  (b)~Sketch of the embedded cluster. Dark (blue) circles: selected atoms of the cobalt leads,
  gray (orange) circles: cobalt atoms in the nanocontact, and empty circles:
  empty spheres around the contact.
  The directions of magnetization in the leads 
  are marked by dark (blue) arrows.
  }
\end{figure}

Our model of the atomic-sized point contact has been built from Co atoms forming two identical
pyramids facing each other between (001) interfaces of fcc Co
as it is shown in Fig.~\ref{fig:contact}(a).
The distance between the central atom and its neighbors was chosen identical
to the fcc nearest neighbor distance, $a$, of 2.506\,\AA.
Note that this geometrical model is the same as the one labelled by C2 in Ref.~\onlinecite{Autes2008},
except that they studied a break-junction between bcc Fe surfaces.
In order to mimic the contraction and
expansion of the contact, the normal to plane distances in the vicinity of the central atom
have been scaled by a factor, hereinafter denoted by $x$, between
0.85 and 1.15, see Fig.~\ref{fig:contact}(a).
A host system assembled of two oppositely magnetized semi-infinite Co leads
and separated by 7 layers of empty spheres (vacuum) is considered.
The embedded cluster in the EC-KKR calculations consisted of 29 ($9+4+1+1+1+4+9$) Co atoms
forming the contact by
substituting empty spheres in the vacuum layers, $16+16$ Co atoms
from the Co surfaces adjacent to the contact, and we also included 80 empty spheres
in the vicinity of the Co atoms in the contact to let the electron density relax around the cluster,
see Fig.~\ref{fig:contact}(b).

First, the electronic structure of the host was calculated in terms of
the fully relativistic screened KKR method applying the surface Green's function
technique.\cite{Szunyogh1994-PRB,Szunyogh1994-JPCM}
 Then the electronic structure of the contact has been determined
 within the EC-KKR method,\cite{Lazarovits2002}
in which the scattering path operator (SPO), corresponding to a finite
cluster, $\mathcal{C}$, embedded into a host system can be obtained from
the following equation,
\begin{equation}
   \label{eq:KKR-EC}
   \bm{\tau}_{\mathcal{C}}(\varepsilon) =
   \left( \mathbf{t}_{\mathcal{C}}^{-1}(\varepsilon) - \mathbf{t}_\text{h}^{-1}(\varepsilon) +
   \bm{\tau}_\text{h}^{-1}(\varepsilon) \right)^{-1},
\end{equation}
where $\mathbf{t}_\text{h}(\varepsilon)$ and $\bm{\tau}_\text{h}(\varepsilon)$
denote the single-site scattering matrix and the SPO matrix for the
host confined to the sites in $\mathcal{C}$, respectively, while
$\mathbf{t}_{\mathcal{C}}$ denotes the single-site scattering matrices of
the embedded atoms. The calculations for both the host and the
cluster were performed within the local spin-density
approximation (LSDA),\cite{Vosko1980} by using the atomic sphere approximation (ASA)
and $\ell_\text{max}=2$ for the angular momentum expansion.

\begin{figure}[!tb]
\includegraphics[width=0.3\columnwidth]{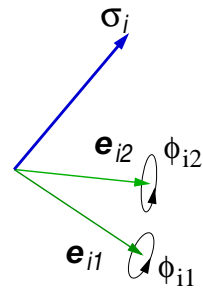}%
\caption{\label{fig:trieder}
(Color online) Sketch of the local frame of reference: the unit vector
$\bm{\sigma}_i$ is parallel to the magnetization at site $i$, while
the unit vectors $\mathbf{e}_{i1}$ and $\mathbf{e}_{i2}$ point into the transverse directions.
Rotations around these axes by $\phi_{i1}$ and $\phi_{i2}$ are also indicated. }
\end{figure}

A fully unconstrained extension of the relativistic EC-KKR method
is used to find the magnetic configuration of the point contact.
The evolution of the atomic magnetic moments
is treated in a semi-classical manner similar to molecular
dynamics, whereby, in spirit of the magnetic force theorem,\cite{Jansen1999}
the driving force is calculated as the derivative of the band energy,
\begin{equation}
  E_\text{b} = \int\limits_{-\infty}^{\varepsilon_\text{F}} \left( \varepsilon-\varepsilon_\text{F} \right) n(\varepsilon)\,\mathrm{d}\varepsilon
             = -\int\limits_{-\infty}^{\varepsilon_\text{F}} N( \varepsilon ) \,\mathrm{d}\varepsilon,
\end{equation}
with respect to the transverse change of the exchange field, where $\varepsilon_\text{F}$ is the Fermi energy,
while $n(\varepsilon)$ and $N( \varepsilon )$ stand for the density of states (DOS) and for the integrated DOS,
respectively.
In the multiple scattering formalism the
exchange field enters the electronic structure via the single-site scattering matrix,
$t_i$.  The first and higher order changes of the $t_i$ matrices as well as the
derivatives of the band energy can straightforwardly be calculated in the local frame of reference
introduced at all sites of the cluster, where the direction vector $\bm{\sigma}_i$
of the magnetization at site $i$, and the two
transverse vectors, $\mathbf{e}_{i1}$ and $\mathbf{e}_{i2}$, form a right handed coordinate
system as shown in Fig.~\ref{fig:trieder}.
The first and second order change of
the single site scattering matrix at site $i$ with respect to rotations by
 $\Delta\phi_{i\alpha}$ around the transverse axes
$\mathbf{e}_{i\alpha}$ can be given by the following commutator formulas,
\begin{align}
  \label{eq:trot1}
  \Delta t_i^{(1)} &= i[\mathbf{e}_{i\alpha}\mathbf{J},t_i]\Delta \phi_{i\alpha} , \\
  \label{eq:trot2}
  \Delta t_i^{(2)} &= -[\mathbf{e}_{i\alpha}\mathbf{J},[\mathbf{e}_{i\beta}\mathbf{J},t_i]] \Delta \phi_{i\alpha} \Delta \phi_{i\beta} ,
\end{align}
where $\mathbf{J}$ is the matrix representation of the total angular momentum operator 
and $\alpha, \beta \in \{1,2\}$. 
Following Ref.~\onlinecite{Udvardi2003}, the first and second derivatives of the
band energy can then be expressed as
\begin{align}
  \label{eq:grad}
    \frac{\partial E_\text{b}}{\partial \phi_{i\alpha}} &= \frac{1}{\pi}\,\mathrm{Re}
    \!\!\int\limits_{-\infty}^{\varepsilon_\text{F}}\!\! \mathrm{Tr}\left \{\tau_{ii}\left
    [\mathbf{e}_{i\alpha}\mathbf{J},m_i\right ]\right \}\,\mathrm{d}\varepsilon ,\\
  \label{eq:hessian}
    \frac{\partial^2 E_\text{b}}{\partial \phi_{i\alpha}\partial \phi_{j\beta}} &=
      - \frac{1}{\pi}\mathrm{Im} \!\!\int\limits_{-\infty}^{\varepsilon_\text{F}}\!\!
      \mathrm{Tr} \left \{\tau_{ij}
      [\mathbf{e}_{j\beta}\mathbf{J},m_j ]
      \tau_{ji}
      [\mathbf{e}_{i\alpha}\mathbf{J},m_i ]\right \}\,\mathrm{d}\varepsilon \nonumber
  \\
    & + \delta_{ij}\frac{1}{\pi}\mathrm{Im} \!\!\int\limits_{-\infty}^{\varepsilon_\text{F}}\!\!
      \mathrm{Tr}\left \{\tau_{ii} [\mathbf{e}_{i\alpha}\mathbf{J},
      [\mathbf{e}_{i\beta}\mathbf{J},m_i ]] \right \}\,\mathrm{d}\varepsilon ,
\end{align}
where $m_i=t_i^{-1}$
and $\tau_{ij}$ is the block of the SPO matrix between sites $i$ and $j$.
Note that for brevity we dropped the energy arguments of the corresponding matrices
in Eqs.~(\ref{eq:trot1}--\ref{eq:hessian}).
In the spirit of a gradient minimization,
rotating the exchange field by a small amount around the torque vector at each sites,
\begin{equation}
  \label{eq:torque}
  \mathbf{T}_i =   \mathbf{e}_{i1} \frac{\partial E_\text{b}}{\partial \phi_{i1}}
                 + \mathbf{e}_{i2} \frac{\partial E_\text{b}}{\partial \phi_{i2}},
\end{equation}
the magnetic configuration gets closer to the local minimum of the energy,  however, the convergence is very slow.
 In order to speed up this procedure, a Newton-Raphson iteration scheme has been applied,
where the inverse of the second derivative tensor, also referred to as the Hessian,
Eq.~(\ref{eq:hessian}), is used to estimate the angle of rotations around the torque
vector given by Eq.~(\ref{eq:torque}). The eigenvalues of the Hessian also provide
information about the stability of the a configuration with zero torque:
if the Hessian is a positive or negative definite matrix then the given configuration
is stable or unstable state of equilibrium, respectively.
Once the Newton-Raphson iteration has converged, new effective potentials and exchange fields
are generated and the procedure is repeated until the effective potential converged and
the torque in Eq.~(\ref{eq:torque}) is decreased below a predefined value of, typically,
$10^{-4}$\,meV.

The starting magnetic configuration for the above optimization procedure
has been determined by Monte Carlo simulated annealing based on a simple isotropic Heisenberg model,
$\mathcal{H} = \frac{1}{2}\sum_{i\neq j} J_{ij}\bm{\sigma}_i\bm{\sigma}_j$,
where $J_{ij}$ is the isotropic exchange coupling between sites $i$ and $j$.
The coupling coefficients between the atomic moments were calculated by using
the torque method proposed by Liechtenstein \textit{et~al}. \cite{Liechtenstein1987}
The exchange couplings have been calculated in a ferromagnetic spin-configuration
parallel to the (100) direction.

In order to avoid the difficulties arising from the continuous degeneracy
of the spin-states in a Heisenberg model, the
magnetization on the central atom was fixed normal to the bulk magnetization.
Considering the inversion symmetry of the point-contact,
only the (1$\overline{1}$0) and the (001) directions
are consistent with the (constrained) magnetic ground-state of the system.
In the first case, the magnetic moments at all sites (layers) remain within the (001) plane,
i.e., normal to the axis of the point-contact,
therefore, in the following this spin-configuration will be termed as a
helical domain wall.
In the second case, all the spin moments are confined to the (1$\overline{1}$0)
plane, thus, we shall call this case the cycloidal domain wall.
Note that the helical and cycloidal spin-configurations closely resemble the Bloch and N\'eel types
of domain walls well-known in bulk and thin-film magnets, respectively.
Since, however, these types of domain walls are distinct through the magnetostatic energy,
to avoid confusion we skipped using the traditional terminology.

\section{Results and discussions}
\subsection{Domain wall configurations}
\label{seq:DW-confs}

\begin{figure}[!tb]
\includegraphics[width=0.6\columnwidth]{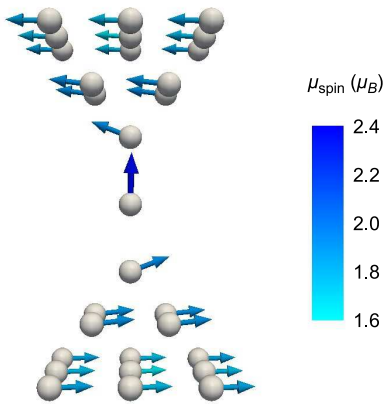}%
\caption{\label{fig:neel_100}
(Color online)
The cycloidal spin-configuration obtained for the unstretched contact ($x=1$).
The lengths of the arrows, indicated also with color coding, are proportional
to the size of the spin magnetic moments.
}
\end{figure}

Self-consistent potentials and exchange fields have been first
determined for both the cycloidal and the helical domain walls and
the Newton-Raphson iterations were started from both initial configurations.
Interestingly, when starting from a helical spin-configuration, the gradients, Eq.~(\ref{eq:grad}),
were initially zero, but the Hessian had a negative eigenvalue 
indicating that the helical spin-configuration belonged to a saddle point of the energy
surface. Throwing the system off this saddle point,
the Newton-Raphson iterations converged to the cycloidal spin-configuration.
Thus, independent of the starting configuration, the magnetic state of the
nanojunction converged to the cycloidal wall structure for the stretching range considered.
In Fig.~\ref{fig:neel_100} the ground-state cycloidal wall configuration is displayed
for $x = 1$.

\begin{figure}[!tb]
\includegraphics[angle=-90,width=\columnwidth]{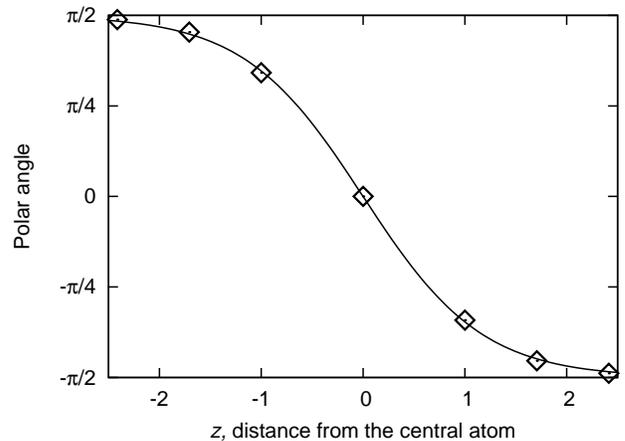}%
\caption{\label{fig:dw} Polar angles averaged within a layer of cobalt atoms
in the contact with $x=1$ as the function of the distance
from the central atom (in units of the fcc nearest neighbor distance,~$a$).
The solid curve displays the fit, $\vartheta(z) = -\frac{\pi}{2} \tanh(2z/d_\text{w})$.
}
\end{figure}

At sites within the same geometrical layer, we obtained fairly similar orientations
for the magnetic moments,
therefore, the shape of the domain wall can well be characterized by orientations
determined as an average within layers. In Fig.~\ref{fig:dw} such a profile
is shown for $x=1$ in terms of polar angles, $\vartheta(z)$.
Remarkably, the well-known analytical form, $\vartheta(z) = -\frac{\pi}{2} \tanh(2z/d_\text{w})$
could be well fitted defining, thus, the width of the domain wall, $d_\text{w}$.
This fit is also shown in Fig.~\ref{fig:dw}.
We note that following Ref.~\onlinecite{Kazantseva2005} the analytical form of a constrained wall profile
should be better described by Jacobian sine functions.
However, testing this alternative approach resulted in to a relative deviation
of less than 0.5~\% in the fitted domain wall thicknesses.

\begin{figure}[!tb]
\includegraphics[angle=-90,width=\columnwidth]{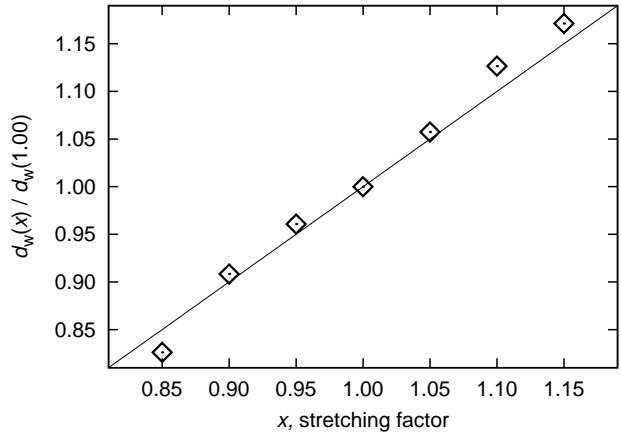}%
\caption{\label{fig:dww} Width of the domain walls through
the point contact as a function of the stretching factor, $x$.
Note that $d_\text{w}(1.00) = 2.34\,a$, where $a$ is the
fcc nearest neighbor distance.
The solid line stands for the identity function.
}
\end{figure}

The change of the width of the domain walls against the length of the point contact
is shown in Fig.~\ref{fig:dww}.
For a clear interpretation, the width of the walls is
normalized to the width of the domain wall for $x=1$.
As obvious from this figure,  $d_\text{w}(x) \approx x \, d_\text{w}(1.00)$  demonstrating that
the width of the domain walls follows the length of the point contact.
In case of Fe$_{20}$Ni$_{80}$ thin films it has been experimentally
found that the constrained geometry can reduce
the width of the N\'eel wall.\cite{Jubert2004} The effect is more pronounced
in ultrathin films of few atomic layers where the width of the domain wall can be as
small as few nanometers in the vicinity of a step edge.\cite{Pietzsch2000}
The effect of the reduced dimensionality is even more obvious in the case of a point contact.
Since the exchange energy gain for the few atoms of the contact
is small compared to the increase of the exchange energy of the leads,
the domain wall can not penetrate into the substrates and the wall is confined to the contact.
The same conclusion has been drawn by
Bruno\cite{Bruno1999} based on a theoretical study of a continuous model of domain walls in a confined
geometry.

\subsection{Magnetic moments}
\label{seq:moments}

\begin{figure}[!tb]
\raisebox{1.33cm}{$\mu_\text{spin}~(\mu_\text{B})$}\hfill\includegraphics[width=0.80\columnwidth]{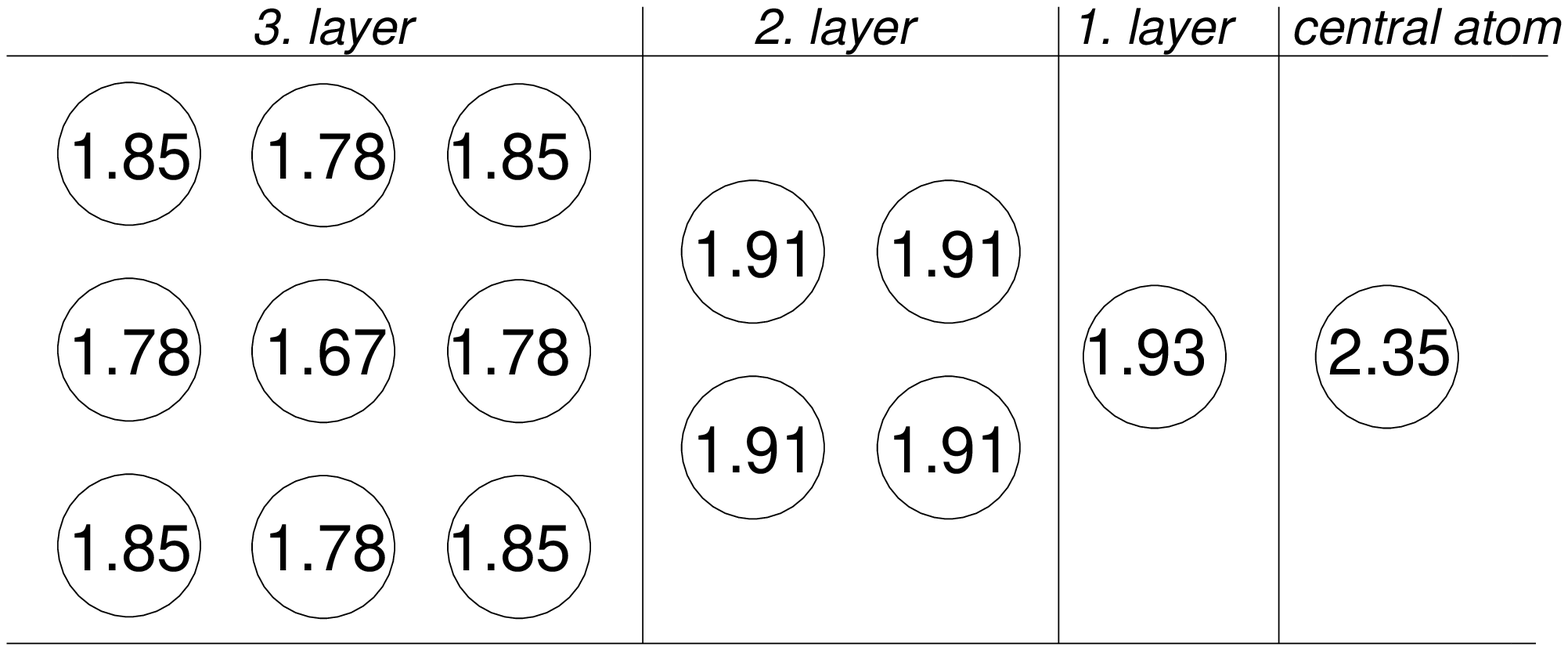}\\%
\raisebox{1.33cm}{$\mu_\text{orb}~ (\mu_\text{B})$}\hfill\includegraphics[width=0.80\columnwidth]{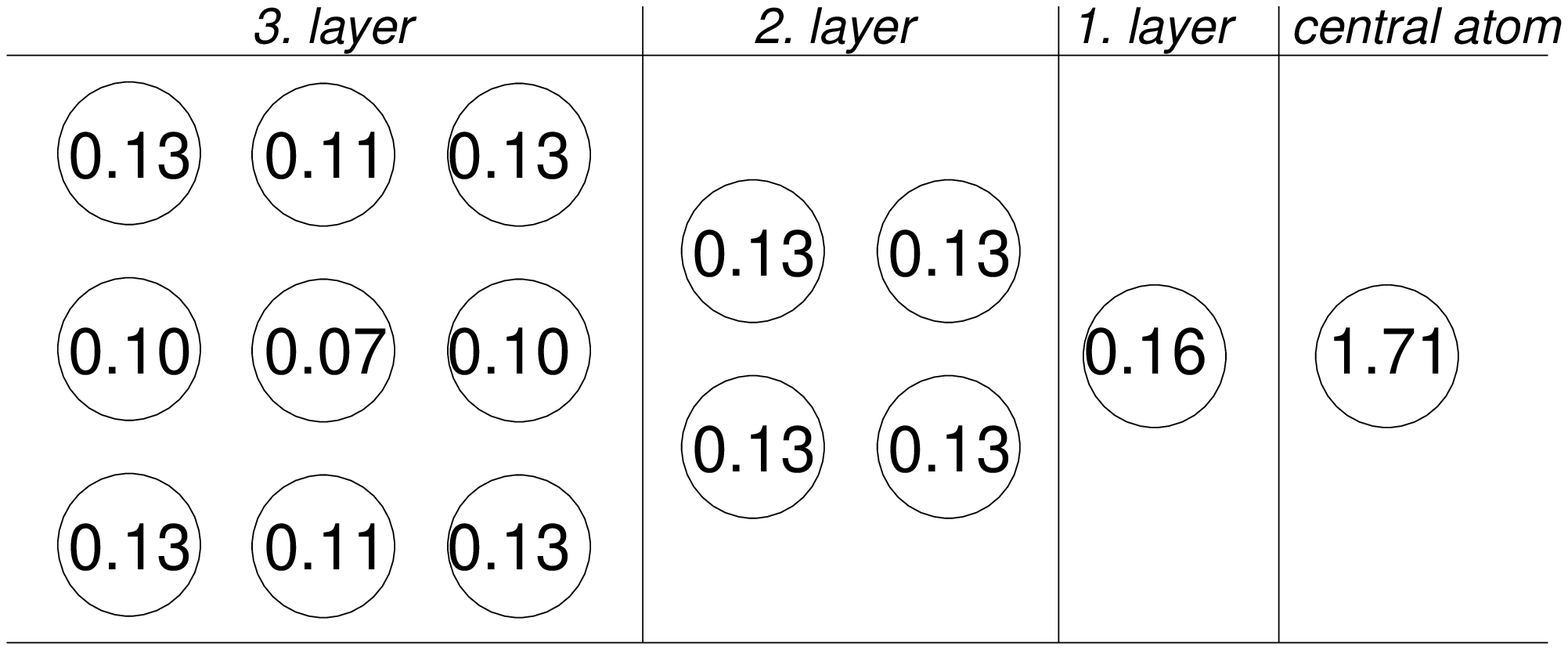}%
\caption{\label{fig:MS}
         \label{fig:MO}
         Calculated atomic magnetic moments ($\mu_B$) in half of the nanocontact
         for the stretching factor, $x=1$.
         In the upper and lower panels shown are the spin and orbital
         moments, $\mu_\text{spin}$ and $\mu_\text{orb}$, respectively.
         For comparison, the spin-moments at the Co surface and in the bulk are
         $1.82\,\mu_\text{B}$ and $1.67\,\mu_\text{B}/\text{atom}$,
         while the corresponding
         values of the orbital moments are $0.14\,\mu_\text{B}$ and $0.08\,\mu_\text{B}$.
}
\end{figure}

The low coordination in thin films and in nanostructures is often accompanied by the
enhancement of the atomic spin and orbital moments.
In Fig.~\ref{fig:MS} the calculated values of the local spin and orbital moments
are given in a point contact with cycloidal wall configuration and stretching factor, $x = 1$.
Since the orbital moment is found almost parallel to the spin-moment at each site,
we presented the projection of the orbital moment to the local spin quantization axis.
Since the contact has a mirror symmetry with respect to the horizontal plane
including the central atom, therefore, the moments in only one half of the contact are displayed.
Our data fit nicely to the
observation reported in Refs.~\onlinecite{Sipr2007} and
\onlinecite{Blonski2009} that the spin and orbital
moments at sites with lower coordination number are larger then at sites with larger
coordination number. This is, in particular, true for the central atom with
coordination number of only two where the values of the spin and orbital moments
are even larger than those obtained for small clusters on Pt(111) and Au(111)
surfaces.\cite{Sipr2007,Blonski2009,Lazarovits2003}

\begin{figure}[!tb]
\includegraphics[angle=-90,width=\columnwidth]{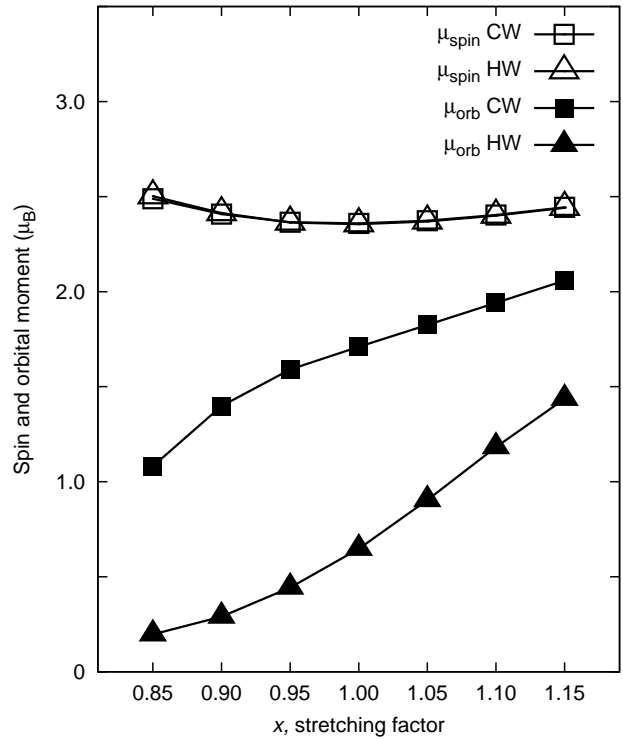}%
\caption{\label{fig:moments}
         The spin- and orbital moments of the central atom
         as a function of the stretching.
         Spin moments are displayed by open symbols, orbital moments are displayed by filled symbols
         as calculated in the cycloidal wall (CW, squares) and in the helical wall (HW, triangles) configurations.
}
\end{figure}

Fig.~\ref{fig:moments} shows the spin and orbital moments of the central atom
as a function of the stretching ratio $x$, for both the cycloidal and the helical
spin-configurations in the point-contact.
Clearly, the spin moments are fairly insensitive to the domain wall configuration:
this can easily be understood as the relative spin-directions are nearly the same
in the two types of domain walls.
Also, there is only a moderate change of the spin moment
in the range of  $2.35\,\mu_\text{B} \le \mu_\text{spin} \le 2.49\,\mu_\text{B}$ for the
stretching ratios under consideration. These values compare well to
$\mu_\text{spin}=2.15\,\mu_\text{B}$ and $\mu_\text{spin}=2.26\,\mu_\text{B}$ calculated for a single Co adatom
on Pt and Au(111) surfaces in Refs.~\onlinecite{Sipr2007}
and \onlinecite{Blonski2009}, respectively.

The dependence of the orbital moment of the central atom on the stretching
is more pronounced than that of the spin moment: in case of a cycloidal and
a helical wall it increases from about $1\,\mu_B$ to $2\,\mu_B$
and from $0.3 \,\mu_B$ to $1.5\,\mu_B$, respectively.
Similar high values of $\mu_\text{orb}$ for the central
atom of a wire-like Fe point-contact were reported in Ref.~\onlinecite{Autes2008}
and attributed to localized atomic-like electronic states treated within
a full Hartree-Fock scheme. It should be mentioned that for a
more reliable description of highly localized states,
the plain LSDA we used in our calculations should be extended with, e.g.,
the local self-interaction correction, LSDA+SIC~\cite{Luders2005}
or the dynamical mean field theory, LSDA+DMFT.~\cite{Kotliar2006}

Apparently, the orbital moment of the central atom is systematically
larger in a cycloidal wall than in a helical wall.
This can be understood since these orbital moments correspond to different
directions: in case of a cycloidal wall it points along the (001) directions,
while, for a helical wall, along the (1$\overline{1}$0) direction.
Such a huge anisotropy of the orbital moment at the central atom
has also been observed in Ref.~\onlinecite{Autes2008}.
According to Bruno's theory~\cite{Bruno1989}
this large orbital momentum anisotropy is related to a large
magnetic anisotropy energy featuring the (001) direction as easy axis,
which clearly corroborates our result for the preference of a cycloidal domain wall.

\subsection{Rotational energy of the domain wall}
\label{seq:MA-DW}

\begin{figure}[!tb]
 \includegraphics[angle=-90,width=\columnwidth]{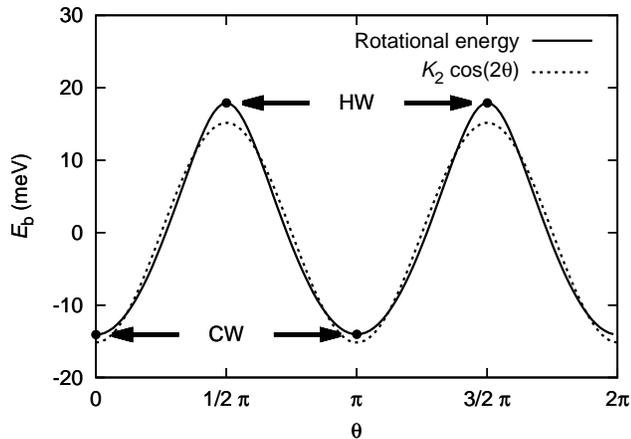}%
 \caption{\label{fig:rotating}
          The band energy of the nanocontact with $x=1.00$ while rotating the exchange field
          at each atomic sites simultaneously around the (110) axis.
          By rotating all the spins by $90^\circ$ the system goes over from the
          cycloidal wall (CW) into the helical wall (HW).
          The dashed line denotes the leading Fourier component of the band energy,
          $-15.2\;[\mathrm{meV}] \cos (2\theta)$, see Eq.~(\ref{eq:EbFourier}).
          Note that we shifted the zero level of the energy to the constant term, $K_0$.}
\end{figure}

The cycloidal and helical spin-configurations of the point contact can be transformed
into each other in term of a simultaneous rotation of the spin-directions
around the axis parallel to the magnetization of the leads.
The energy along the path of this global rotation, termed as the rotational
energy of the domain wall, was calculated using the magnetic
force theorem, namely, from the band energy of the system by rotating the orientation
of the exchange field at each atomic site around
the (110) axis and keeping frozen the effective potentials and fields as obtained for the ground
state cycloidal wall configuration.
For the case of the unstretched configuration the results are plotted in Fig.~\ref{fig:rotating}.
The two minima and maxima of the band energy belong to the two-fold degenerate
cycloidal and helical domain wall configurations.
The height of the energy barrier between the two ground state cycloidal spin-configurations is 32.0\,meV.
Similar behavior has been found for the whole stretching range of the point contact.
The energy differences between the two types of domain wall as a function of the
stretching ratio are displayed by diamonds in Fig.~\ref{fig:difference}.

Due to time reversal symmetry, the magnetic anisotropy energy has a periodicity of $\pi$,
but it does not comply with a usual $\cos^2(\theta)$ dependence.
To explore this deviation we performed the Fourier expansion,
\begin{equation}
E_\text{b}(\theta) = K_0 + \sum_{k=2,4,\dots}^{\infty}{K_{k} \cos(k\theta)} \, ,
\label{eq:EbFourier}
\end{equation}
for the contacts with different stretching. Note that because of the inversion
symmetry of the contact $E_\text{b}(\theta)=E_\text{b}(\pi-\theta)$ applies,
therefore, the $\sin(k\theta)$ ($k=2,4,\dots)$ terms do not appear in
the expansion, Eq.~(\ref{eq:EbFourier}).  We summarized the Fourier
coefficients, $K_k$, in Table~\ref{tab:Fouriercomponents}.
We found that in each case the term $K_2\cos(2\theta)$ adds the largest weight
to the rotational energy of the domain wall.
The next term $K_4\cos(4\theta)$ is quite significant for $x \ge 0.95$,
but it drops for smaller stretching.  Interestingly, in the stretching range of $x \le 0.90$
the $k=6$ term overweights the one of $k=4$,
whereas in the complementary range the $k=6$ term is negligible.
It should be mentioned that the $k\ge8$ terms of the Fourier
expansion have practically vanishing weight.

\begin{table}[!tb]
  \centering
  \caption{\label{tab:Fouriercomponents}
           The $k=2$, $4$ and $6$ Fourier coefficients (in units of meV)
           of the rotational energy of the point-contact, Eq.~(\ref{eq:EbFourier}),
           as a function of the stretching parameter, $x$.
  }
  \begin{ruledtabular}
  \begin{tabular}{dddd}
    \multicolumn{1}{c}{$x$}
      & \multicolumn{1}{c}{$k=2$}
      & \multicolumn{1}{c}{$k=4$}
      & \multicolumn{1}{c}{$k=6$} \\
  %  x  &    k=2  &    k=4  &    k=6 \\
  \hline 
  0.85  &   -6.3  &   0.15  &  0.397 \\
  0.90  &  -10.0  &   0.36  &  0.499 \\
  0.95  &  -13.6  &   1.40  &  0.298 \\
  1.00  &  -15.2  &   2.17  & -0.040 \\
  1.05  &  -15.1  &   2.35  & -0.122 \\
  1.10  &  -14.4  &   2.24  & -0.083 \\
  1.15  &  -13.2  &   1.92  &  0.025 \\
  \end{tabular}
  \end{ruledtabular}
\end{table}

\subsection{Magnetic anisotropy of the central atom}
\label{seq:MA-CA}

As we have seen in Sec.~\ref{seq:moments}, the central atom of the contact
exhibits a huge orbital moment anisotropy that should be accompanied by 
a large magnetic anisotropy energy. For that reason, we analyze the
band energy of the point-contact, $E_\text{b}(\bm{\sigma})$,
with $\bm{\sigma}$ denoting the spin-orientation at the central atom,
whereas the spin-orientations of all the other sites in the contact
are kept fixed as obtained in the ground-state cycloidal wall configuration.

Our analysis is based on an expansion of $E_\text{b}(\bm{\sigma})$
in terms of (real) spherical harmonics, $R_\ell^m(\bm{\sigma})$,
\begin{equation}
E_\text{b}(\bm{\sigma})
= \sum_{\ell,m} K_\ell^m R_\ell^m(\bm{\sigma}) \, ,
\label{eq:Eanis}
\end{equation}
with the angular momentum indices, $\ell=0,1,2,\dots$ and $-\ell \le m \le \ell$.
Similar to the rotational energy of the domain wall, we used the magnetic force theorem
to evaluate $E_\text{b}(\bm{\sigma})$,
but here we employed Lloyd's formula,~\cite{Lloyd1967} since it accurately
accounts for the change of the band energy of the whole point-contact
with respect to the change of the spin-orientation at the central site.
For the expansion, the integration over $\bm{\sigma}$ was performed using a 51 points
Gaussian quadrature along the $z$-direction and a uniform mesh of 100 points in the azimuth angle,
resulting in a spherical grid of 5100 points.
The obtained coefficients are summarized in Table~\ref{tab:anis} up to $\ell=4$ and for all the
stretching ratios under consideration. Only the non-vanishing coefficients are presented,
for clarity, together with the definition of the corresponding spherical harmonics,
$R_\ell^m(\bm{\sigma})$.

\begin{table*}[!tb]
  \caption{\label{tab:anis}
           Expansion coefficients $K_\ell^m$ (in units of meV) of the band energy of the contact,
           see Eq.~(\ref{eq:Eanis}), according to real spherical harmonics $R_\ell^m$ up to $\ell=4$.
           }
  \begin{ruledtabular}
  \begin{tabular}{cccddddddd}
  $\ell$ & $m$ & $R_\ell^m \vphantom{\sqrt\frac12}$                               &  x=0.85  &  x=0.90  &  x=0.95  &  x=1.00  &  x=1.05  &  x=1.10  &  x=1.15  \\
  \hline
  1 & 0 &               $\frac12 \sqrt{\frac3\pi} z$                              & -240     & -247     & -235     & -212     & -192     & -176     & -159     \\
  \hline
  2 & 0 &      $\frac14 \sqrt{\frac{5}{\pi}}  \left(3z^2-1\right)$                &  -25.3   &  -30.0   &  -33.2   &  -32.4   &  -30.9   &  -28.4   &  -25.6   \\
  2 & 2 &      $\frac14 \sqrt{\frac{15}{\pi}} \left(x^2-y^2\right)$               &    4.30  &    2.54  &    1.39  &    0.51  &   -0.29  &   -0.92  &   -1.36  \\
  \hline
  3 & 0 &      $\frac14 \sqrt{\frac{7}{\pi}}  \left(5z^3-3z\right)$               &    4.12  &    3.06  &    1.63  &    0.71  &   -0.28  &   -1.43  &   -2.67  \\
  3 & 2 &      $\frac14 \sqrt{\frac{105}{\pi}}\left(x^2-y^2\right)z$              &   -0.199 &   -0.093 &    0.004 &    0.108 &    0.196 &    0.267 &    0.293 \\
  \hline
  4 & 0 & $\frac{3}{16} \sqrt{\frac1\pi} \left(35z^4-30z^2+3\right)$              &   -0.63  &    1.72  &    4.60  &    4.94  &    5.05  &    4.85  &    4.32  \\
  4 & 2 & $\frac38 \sqrt{\frac5\pi}      \left(x^2-y^2\right)\left(7z^2-1\right)$ &    0.033 &    0.125 &    0.184 &    0.108 &    0.051 &    0.001 &   -0.052 \\
  4 & 4 & $\frac{3}{16} \sqrt{\frac{35}{\pi}} \left(x^4-6x^2y^2+y^4\right)$       &   -0.007 &   -0.005 &   -0.018 &   -0.041 &   -0.088 &   -0.187 &   -0.345 \\
  \end{tabular}
  \end{ruledtabular}
\end{table*}

The absence of certain spherical harmonics in expansion Eq.~(\ref{eq:Eanis}) can be discussed
based on group-theoretical arguments. The function $E_\text{b}(\bm{\sigma})$ should be
invariant under symmetry transformations, $g$, of the point-contact,
$E_\text{b}(\bm{\sigma}) = E_\text{b}(g\bm{\sigma})$, including the symmetry of both
the lattice and the given (cycloidal) spin-configuration. Regarding that the
spin-vectors transform as axial vectors, the only allowed transformation is the
reflection onto the (001) plane: $(x,y,z) \rightarrow (-x,-y,z)$.
Thus we conclude that only those function can enter the expansion of $E_\text{b}(\bm{\sigma})$
that contain even powers of the variables $x$ and $y$.
As seen from Table~\ref{tab:anis}, this is fully confirmed by our calculations.
Apparently, the expansion Eq.~(\ref{eq:Eanis}) shows a satisfactory convergence
as the coefficients rapidly decrease with increasing $\ell$.
An obvious exception can, however, be seen for $K_4^0$ that for $x \ge 0.95$ overweights $K_3^0$.
Noticeably, among the terms with a given $\ell$, the one associated with the $z$ component
of the magnetization ($m=0$), i.e., excluding in-plane anisotropy, has the largest weight.

In order to connect the above results to the rotational energy of the domain wall discussed
in Sec.~\ref{seq:MA-DW}, we relate expansion Eq.~(\ref{eq:Eanis}) to
a classical spin-model.  According to a Heisenberg model extended by
relativistic corrections~\cite{Udvardi2003,Szunyogh2011} the energy in Eq.~(\ref{eq:Eanis})
can be expressed as
\begin{equation}
  \label{eq:E-spinmodel}
  E(\bm{\sigma}) =   E_\text{anis}(\bm{\sigma}) + \bm{\sigma}\sum_j \mathbf{J}_{\text{c}j} \bm{\sigma}_j \, ,
\end{equation}
where $\mathbf{J}_{\text{c}j}$ denote the exchange coupling tensor between the central site and the other
sites of the contact with classical spin-vectors $\bm{\sigma}_j$
and $E_\text{anis}(\bm{\sigma})$ stands for the on-site anisotropy energy that,
due to the tetragonal ($D_{4h}$) point-group symmetry of the point-contact, can be expanded
up to $\ell=4$ as
\begin{equation}
\label{eq:anis-tetragonal}
E_\text{anis}(\bm{\sigma})=
K_2^0 R_2^0(\bm{\sigma}) + K_4^0 R_4^0(\bm{\sigma}) +  K_4^4 R_4^4(\bm{\sigma}) \, .
\end{equation}

It is clear that the $(\ell,m)=(1,0)$ term in Eq.~(\ref{eq:Eanis}) is uniquely related to the
exchange coupling and, due to the presence of a cycloidal wall, it represents a
strong Weiss field that orients the magnetic moment at the central site
along the $z$ direction. Because of the increasing distances between the central
site and the other sites of the contact, it is also easy to understand why this
term significantly decreases with increasing stretching ratio. On the other hand, 
there is no $(\ell,m)=(1,0)$ term in the rotational energy 
of the domain wall, Eq.~(\ref{eq:EbFourier}),
since in that case the relative orientation of the spins are unchanged.
With other words, repeating the expansion Eq.~(\ref{eq:Eanis})
in the presence of a helical wall, the leading term correspond to the
spherical harmonics $\propto x$, with practically the same coefficients as listed
in Table~\ref{tab:anis} for $(\ell,m)=(1,0)$.

In relation to Eq.~(\ref{eq:anis-tetragonal}), the terms proportional to $R_2^0$,
$R_4^0$ and $R_4^4$ in Eq.~(\ref{eq:Eanis}) can mainly be attributed to
on-site anisotropy contributions to the spin-Hamiltonian, however, the effect
of higher order spin-interactions can not be ruled out.
The second-order uniaxial anisotropy coefficients, $K_2^0$, are negative
in the whole range of stretching, favoring thus a normal-to-plane direction.
Remarkably, the magnitude of $K_2^0$ is around 30\,meV,
with a maximum of $\left|K_2^0\right|=33.2\,\mathrm{meV}$ at $x = 0.95$.
This value should be compared to some results communicated in the literature:
Etz \etal\cite{Etz2008} and Bornemann \etal
\cite{Bornemann2007} calculated 5.3\,meV and 4.76\,meV, respectively, for the MAE of
a Co ad-atom on Pt(111) surface, while, including orbital polarization,
Gambardella \etal \cite{Gambardella2003} obtained
18.45\,meV for the same system.
In a similar geometrical confinement of an  atomic scale junction,
W and Ir turned out to be magnetic with a magnetic anisotropy energy
comparable to our values.\cite{Thiess2010}

\begin{figure}[!tb]
 \includegraphics[angle=-90,width=\columnwidth]{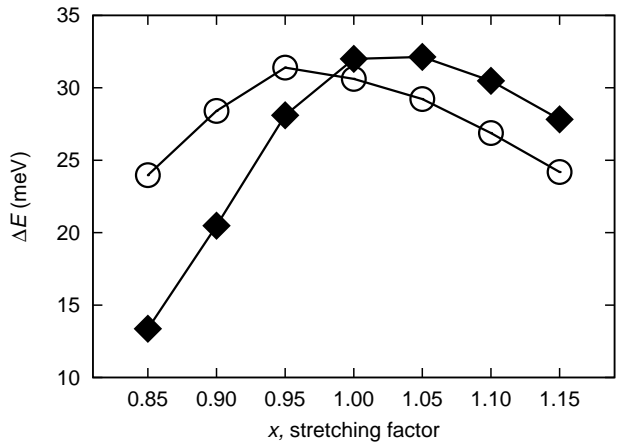}
 \caption{\label{fig:difference}
          Diamonds: Calculated energy differences between the helical and cycloidal domain walls,
          $E_\text{HW}-E_\text{CW}$,
          circles: on-site uniaxial magnetic anisotropy energy of the central atom (see text)
          as a function of the stretching parameter,~$x$.
          Thin lines serve as a guide for the eye.
}
\end{figure}

From Fig.~\ref{fig:rotating} and Table~\ref{tab:Fouriercomponents} we inferred that the rotational energy of the
domain wall is dominated by the uniaxial magnetic anisotropy term
proportional to $\cos^2\theta = z^2$.
In Fig.~\ref{fig:difference} the energy differences obtained between the
helical wall configuration and the ground state cycloidal wall configuration
is plotted as a function of the stretching factor, together with that
provided by the uniaxial anisotropy of the central atom,
$\frac{3}{4}\sqrt{\frac{5}{\pi}} K_2^0$.
The values of $\Delta E$ from the two calculations agree well for $x \ge 0.95$,
while for more squeezed contacts
the uniaxial anisotropy of the central atom overestimates
the energy difference between the different types of domain walls.
Nevertheless, we can in general conclude that the main driving force of
the formation of a cycloidal domain wall is a giant uniaxial on-site magnetic anisotropy
at the central atom: in the cycloidal wall the magnetic moment of the central atom
is parallel to the easy axis, while in the helical wall configuration it lies
within the hard plane.

Finally, we briefly comment on the terms corresponding to $(\ell,m)=(2,2)$,
$(3,0)$, $(3,2)$ and $(4,2)$ in Table~\ref{tab:anis}.
Since these terms are not invariant under transformations of the $D_{4h}$ point-group,
they can not be accounted for the on-site anisotropy terms.
In terms of a spin-model, these terms should, therefore, be related to higher order
spin-interactions. The $(\ell,m)=(2,2)$ term can, e.g., be identified as
the consequence of biquadratic interactions,\cite{Deak2011}
$\sum_{i} B_{\text{c}i} (\bm{\sigma}\bm{\sigma_i})^2$, while the $\ell=3$ terms to
triquadratic interactions,\cite{Boca2012}  $\sum_{i} T_{\text{c}i} (\bm{\sigma}\bm{\sigma_i})^3$.
Four-spin interactions have been explicitely calculated and proved to give
significant contributions to a spin-Hamiltonian of Cr trimers deposited
on Au(111) surface by Antal \textit{et~al.},\cite{Antal2008} but recently their
presence was highlighted even in bulk magnets.~\cite{Lounis2010}

\section{Summary}

In case of deposited magnetic nanostructures the point-group symmetry of the system might
considerably be reduced, therefore, complex magnetic states occur naturally.
Detecting and investigating such magnetic states pose a challenge for ab initio calculations.
We have developed a computational technique based on a self-consistent embedded cluster Korringa-Kohn-Rostoker method
suitable to find non-collinear ground-states of finite magnetic clusters.
The method is applied to determine the structure of a domain wall
formed through an atomic scale nanocontact between two antiparallelly magnetized cobalt leads.
The obtained ground state is a cycloidal domain wall which remains stable
against squeezing or stretching the contact along the normal-to-plane direction.
A huge enhancement, as well as, anisotropy of the orbital moment are found at the central site
of the contact.
The energy of the domain walls was explored in terms of the magnetic force theorem.
Our main observation is that the formation of the cycloidal wall against a helical wall is
primarily driven by the uniaxial on-site anisotropy at the central site.
We also found effects of higher order spin-interactions as terms
in the expansion of the band energy not complying with the point-group symmetry of the point contact. 

\begin{acknowledgments}
Financial support is acknowledged to
the New Sz\'echenyi Plan of Hungary Project ID.~T\'AMOP-4.2.2.B-10/1--2010-0009,
the Hungarian Scientific Research Fund (contracts OTKA PD83353, K77771, K84078) and
the Bolyai Research Grant of the Hungarian Academy of Sciences.
UN and LS acknowledge financial support from the DFG through SFB 767.
\end{acknowledgments}

\bibliography{co_contact}

\end{document}